\documentclass[%
 aip,
 amsmath,amssymb,
 reprint,%
]{revtex4-1}

\usepackage{graphicx}
\usepackage{dcolumn}
\usepackage{bm}

\usepackage[utf8]{inputenc}
\usepackage[T1]{fontenc}
\usepackage{mathptmx}
\usepackage{etoolbox}

\usepackage[unicode]{hyperref}
\hypersetup{
	unicode=true, 
	plainpages=false,
	colorlinks=true,
	linkcolor=blue,
	citecolor=blue,
	filecolor=blue,
	urlcolor=blue
}
\usepackage{filemod}

\newcommand{\ket}[1]{|#1\rangle}

\newcommand{\ChM}[1]{\textcolor{black}{#1}}

\newcommand{\boldvec}[1]{\ensuremath{\boldsymbol{\mathbf{#1}}}}
\newcommand{\spvec}[1]{\ensuremath{\boldvec{#1}}}
\newcommand{\unitvec}[1]{\ensuremath{\hat{\boldvec{#1}}}}

\newcommand{\customsub}[1]{\ensuremath{_{#1}}}

\newcommand{\pol}{\unitvec{e}}
\newcommand{\rv}{{\spvec{r}}}

\makeatletter
\def\@email#1#2{%
 \endgroup
 \patchcmd{\titleblock@produce}
  {\frontmatter@RRAPformat}
  {\frontmatter@RRAPformat{\produce@RRAP{*#1\href{mailto:#2}{#2}}}\frontmatter@RRAPformat}
  {}{}
}%
\makeatother
\begin{document}

\preprint{AIP/123-QED}

\title[Topological optical skyrmion transfer to matter]{Topological optical skyrmion transfer to matter}
\author{Chirantan Mitra}
\affiliation{Nanyang Quantum Hub, School of Physical and Mathematical Sciences, Nanyang Technological University, 21 Nanyang Link, Singapore 637371, Singapore}
\affiliation{MajuLab, International Joint Research Unit IRL 3654, CNRS, Universit\'e C\^ote d'Azur, Sorbonne Universit\'e, National University of Singapore, Nanyang
Technological University, Singapore}
\author{Chetan Sriram Madasu}
\affiliation{Centre for Quantum Technologies, National University of Singapore, 117543 Singapore, Singapore}
\affiliation{MajuLab, International Joint Research Unit IRL 3654, CNRS, Universit\'e C\^ote d'Azur, Sorbonne Universit\'e, National University of Singapore, Nanyang
Technological University, Singapore}
\author{Lucas Gabardos}
\affiliation{Nanyang Quantum Hub, School of Physical and Mathematical Sciences, Nanyang Technological University, 21 Nanyang Link, Singapore 637371, Singapore}
\affiliation{MajuLab, International Joint Research Unit IRL 3654, CNRS, Universit\'e C\^ote d'Azur, Sorbonne Universit\'e, National University of Singapore, Nanyang
Technological University, Singapore}
\author{Chang Chi Kwong}
\affiliation{Nanyang Quantum Hub, School of Physical and Mathematical Sciences, Nanyang Technological University, 21 Nanyang Link, Singapore 637371, Singapore}
\affiliation{MajuLab, International Joint Research Unit IRL 3654, CNRS, Universit\'e C\^ote d'Azur, Sorbonne Universit\'e, National University of Singapore, Nanyang
Technological University, Singapore}
\author{Yijie Shen}
\affiliation{Centre for Disruptive Photonic Technologies, School of Physical and Mathematical Sciences, Nanyang Technological University, 21 Nanyang Link, Singapore 637371, Singapore}
\affiliation{School of Electrical and Electronic Engineering, Nanyang Technological University, Singapore 639798, Singapore}
\author{\ChM{Janne Ruostekoski}}
\affiliation{Department of Physics, Lancaster University, Lancaster, LA1 4YB, United Kingdom}
\author{\ChM{David Wilkowski}}
\affiliation{Nanyang Quantum Hub, School of Physical and Mathematical Sciences, Nanyang Technological University, 21 Nanyang Link, Singapore 637371, Singapore}
\affiliation{MajuLab, International Joint Research Unit IRL 3654, CNRS, Universit\'e C\^ote d'Azur, Sorbonne Universit\'e, National University of Singapore, Nanyang
Technological University, Singapore}
\affiliation{Centre for Quantum Technologies, National University of Singapore, 117543 Singapore, Singapore}
\affiliation{Centre for Disruptive Photonic Technologies, School of Physical and Mathematical Sciences, Nanyang Technological University, 21 Nanyang Link, Singapore 637371, Singapore}

\date{\today}

\begin{abstract} 
The ability of structured light to mimic exotic topological skyrmion textures, encountered in high-energy physics, cosmology, magnetic materials, and superfluids has recently received considerable attention. Despite their promise as mechanisms for data encoding and storage, there has been a lack of studies addressing the transfer and storage of the topology of optical skyrmions to matter. Here, we demonstrate a high-fidelity mapping of skyrmion topology from a laser beam onto a gas of cold atoms, where it is detected in its new non-propagating form. Within the spatial overlap of the beam and atom cloud, the skyrmion topological charge is preserved, with a reduction from $Q \simeq 0.91$ to $Q \simeq 0.84$ mainly due to the beam width exceeding the sample size. Our work potentially opens novel avenues for topological photonics state storage and the analysis of more complex structured light topologies.
\end{abstract}

 \maketitle

\section{Introduction}

Structured light~\cite{strlight_review} of spatially varying polarization patterns, as a field of topological photonics, has recently garnered significant interest. In paraxial optics, the polarization of a light beam is commonly described by Stokes vectors. A monotonically varying Stokes vector, covering all orientations over the beam's cross-sectional area, can be represented as a two-dimensional skyrmion texture~\cite{shen2024optical,Gao2020,Donati2016}. These `baby-skyrmions' are non-singular, localized topological objects that take a uniform asymptotic value sufficiently far from the center of the structure. They are lower-dimensional analogs of three-dimensional skyrmions, originally introduced in nuclear physics~\cite{Skyrme1961, manton-sutcliffe, Battye97,Battye09} and later studied in superfluids~\cite{Volovik1977,Shankar1977, Ruostekoski2001, al-khawaja_nature_2001, battye_prl_2002,savage_prl_2003, Lee2018}, elementary particle physics and cosmology~\cite{radu_physrep_2008}. Owing to their easier experimental accessibility and recognized applications, two-dimensional skyrmions are generally better known. In superfluid liquid $^3$He, they are also known as Anderson-Toulouse-Chechetkin~\cite{anderson_prl_1977, chechetkin_jetp_1976} and Mermin-Ho~\cite{mermin_prl_1976} vortices. In magnetic materials~\cite{muhlbauer_science_2009,Nagaosa2013}, they hold great promise for data storage and spintronics, and they have been prepared in exciton-polariton structures~\cite{Cilibrizzi2016, Donati2016, Krol2021, Lorenzo2021}. Analogous skyrmions have also been theoretically studied~\cite{ho_prl_1998,Ohmi1998,mizushima_prl_2002,Lovegrove2014} and experimentally observed~\cite{leanhardt_prl_2003,choi_prl_2012,weiss_ncomm_2019} in atomic Bose-Einstein condensates.

Although skyrmions in optical fields~\cite{shen2024optical} are a more recent subject, they have already been intensively studied theoretically~\cite{Gao2020,Gutierrez-Cuevas2021,Sugic2021,Shen21,Parmee2022,Liu2022,cisowski2023building,McWilliam23,Ye24} and demonstrated in laboratory experiments, both in evanescent fields~\cite{Tsesses2018, Du2019,Davis2020,dai2020plasmonic,Zhang24} and in free space~\cite{Sugic2021,shen2021supertoroidal,shen2022generation,Ornelas24}.
Skyrmions are valuable for exploring fundamental aspects of topological photonics and structured light and hold potential applications in metrology and imaging techniques~\cite{Du2019,Davis2020}, as well as in optical communication and information storage~\cite{Wan2023,shen2024optical}. A key element for many of these applications is the high-fidelity mapping of skyrmion topology from optical fields to matter, its storage, processing, and retrieval. However, experimental studies addressing these processes have been notably absent. In fact, there has been an almost complete lack of even reports on the interaction of optical skyrmions with matter, with the notable exception of a recent study addressing the imprinting of surface relief structures of polarization patterns on an azopolymer film~\cite{Tamura24}. This method, however, only imprints rippled marks of the polarization structures on the material surface without transferring the topological properties of the skyrmion.

In this work, we prepare a topological skyrmion texture in a laser beam and measure its topological charge. This optical skyrmion is then made to interact with atomic matter. By leveraging the well-established technology of imprinting optical vortices in Laguerre-Gaussian (LG) laser beams onto atomic vortices in atom clouds~\cite{Andersen06,Pugatch07,Wright09,Moretti09}, we demonstrate a high-fidelity mapping of skyrmion topology onto matter, enabling it to be detected in its new form of an atomic pseudospin excitation. Transfer of skyrmions and hopfions from light to atoms was recently analyzed theoretically in Ref.~\cite{Parmee2022} via internal level transitions. In our approach, we take an alternative route, achieving state transfer through adiabatic passage in a dark state. This mechanism is analogous to electromagnetically-induced transparency, which has been successfully employed to slow and store light~\cite{LiuEtAlNature2001,lukin2003colloquium,FleischhauerEtAlRMP2005}. The topological charge density within the matter is determined by detecting the atomic dark state population and projecting onto the diabatic state. By integrating the measured charge density, we obtain a topological charge of $Q \simeq 0.84$ in the atomic case. This value is slightly reduced compared to the original $Q \simeq 0.91$ observed in the optical skyrmion, primarily due to the chosen laser beam width being significantly larger than the atom cloud. Within the spatial overlap region, the integrated topological charge is effectively transferred, while the different coupling strengths associated with the light polarization lead to a slight spatial stretching of the skyrmion texture.

\section{Results}

Mathematically, we represent a baby-skyrmion as a non-singular topological texture on a plane determined by a pseudospin. This pseudospin assumes a uniform constant value along some closed path that encloses the origin and, for a full integer-winding skyrmion, points in every direction somewhere within the region enclosed by the path. The entire region defined by the path can be considered analogous to a unit sphere $S^2$, with the spin directions characterized by topologically non-trivial $S^2\rightarrow S^2$ mappings. The topological charge, which counts how many times the object wraps over $S^2$, is given by
\begin{equation}\label{Eq:SkyrmionNumber}
Q =\int_\mathcal{S}\frac{d\Omega_i}{8\pi} \epsilon_{ijk} \textbf{S}\cdot\frac{\partial\textbf{S}}{\partial r_j}\times \frac{\partial \textbf{S}}{\partial r_k},
\end{equation}
where ${\bf S}$ is the spin vector at spatial coordinates ${\bf r}$, $\epsilon_{ijk}$ represents the completely antisymmetric Levi-Civita tensor, and $d\Omega_i$ is an area element of a surface $\mathcal{S}$ that covers the texture.

For a light field ${\bf E}=[E\customsub{R},E\customsub{L}]^T$, with two normalized transverse polarization components $E\customsub{R}$ and $E\customsub{L}$, this parametrization is commonly employed by the Stokes vector ${\bf S}$~\cite{BOR99}. The Stokes vector is defined by its components  $ S_j= {\bf E}^\dagger \sigma_j {\bf E}$ in terms of the Pauli matrices $\sigma_j$, and it takes values on the $S^2$ Poincar\'e sphere. Field configurations that satisfy $Q=1$ --- indicating a Stokes vector that wraps over the Poincar\'e sphere exactly once --- correspond to Poincar\'e beams~\cite{Beckley2010} that can be generated by a superposition of a Gaussian and an LG beam~\cite{Donati2016,Gao2020}
\begin{equation}\label{eq:light}
    {\bf E}(\textbf{r}) = U_{0,0}(\textbf{r}) \pol\customsub{R} + U_{1,0}(\textbf{r}) \pol\customsub{L}.
\end{equation}
Here $U_{l,p}(\textbf{r}) $ are the LG modes with azimuthal quantum number $l$ and radial quantum number $p$. 

\begin{figure}[t!]
 \centering
 \includegraphics[width=0.45\textwidth]{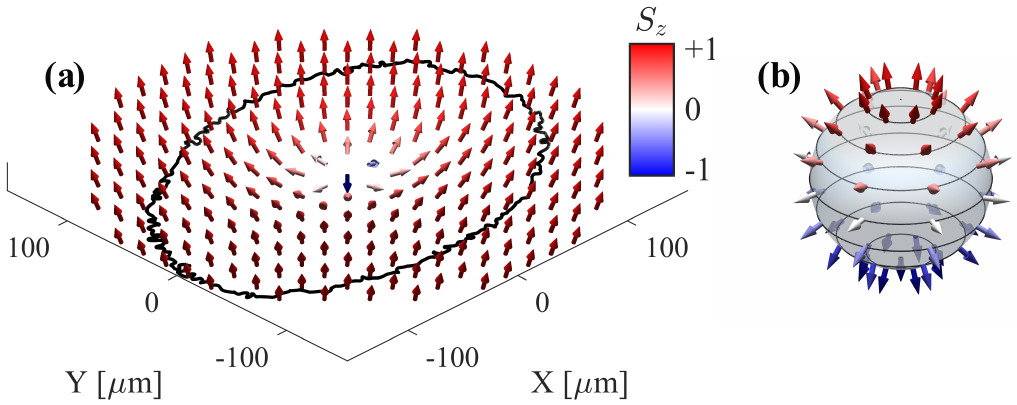}
\caption{\textbf{Experimentally measured topological optical skyrmion texture in a laser beam.} \textbf{(a)} The skyrmion texture represented by the Stokes vectors on the cross-sectional plane of the laser beam. The solid-black curve determines the path along which the Stokes vector maintains a constant angle 0.88$\pi$ between the radial and vertical axes. \textbf{(b)} The region outlined by this path is mapped to a sphere to determine the $S^2\rightarrow S^2$ mapping of the skyrmion topology. The structure within the outlined region corresponds to a topological charge value of $Q \simeq 0.91$.}

 \label{fig:optical_Skyrmion_texture}
\end{figure}

In Fig.~\ref{fig:optical_Skyrmion_texture}, we present the experimentally measured pseudospin profile of the Stokes vectors for the propagating light field along the beam's transverse direction before interaction with the atomic sample. 
Since our focus is on transferring the topological properties of the skyrmion to matter rather than on skyrmion light propagation, we did not attempt to demonstrate the stability of the solution \eqref{eq:light} during propagation~\cite{shen2024optical}.
The optical skyrmion topology is depicted by illustrating the Stokes vectors both on a 2D plane at a cross-section of the laser field and on a sphere. Compactification on a sphere is achieved by defining a closed path on the plane along which the Stokes vector maintains a constant angle between the radial and vertical ($z$) axes. From the experimental data, we determine the largest such path that allows integration of the skyrmion topological charge and find that it corresponds to an angle of 0.88$\pi$ (see solid-black curve in Fig.~\ref{fig:optical_Skyrmion_texture}a). The region outlined by this path then serves to determine a compactification region for the skyrmion topology and is projected onto the sphere, with the points along the path representing the north pole. The Stokes vector wraps over the Poincar\'e sphere almost entirely once, giving a topological charge of $Q \simeq 0.91$ in Eq.~\eqref{Eq:SkyrmionNumber}.

Here, we demonstrate how this topological skyrmion in a light field is mapped onto an atomic excitation, and subsequently detected in this new form. We represent the topological skyrmion texture in matter in terms of the pseudospin vector ${\bf S} = \Psi^\dagger {\boldsymbol \sigma}  \Psi$, defined by the normalized spinor $\Psi=[\psi_1, \psi_2]^T$ of two atomic components, $\psi_1$ and $\psi_2$. The spinor is described by a complex ratio ${\psi_2}/{\psi_1} = \tan(\beta/2) e^{i \alpha}$, or in terms of the pseudospin components
\begin{equation}
    \psi_1=  \cos(\beta/2) e^{-i \alpha/2},\quad
\psi_2=  \sin(\beta/2) e^{i \alpha/2}.
\label{eq:skystate}
\end{equation}
To simplify the analysis, we initially consider an infinite $xy$ plane of an atomic gas. In this setup, the spherical angles $(0\leq\beta\leq\pi,0\leq\alpha\leq2\pi)$ describe a Poincar\'e sphere that is wrapped over exactly once when we identify a mapping between the Poincar\'e sphere coordinates and the polar coordinates on the plane, where the radius $\rho= \tan(\beta/2)$ and the azimuthal angle $\varphi=\alpha$. The non-trivial topology of this configuration is captured by writing $Q=\int dx dy {\cal Q}(\rv)$ in Eq.~\eqref{Eq:SkyrmionNumber} in terms of the topological charge density ${\cal Q}(\rv)$. For the unit vector $\hat{\bf z}$ normal to the plane, this parametrization yields ${\cal Q}=\hat{\bf z}\cdot (\nabla\alpha\times \nabla\cos\beta)$~\cite{Sugic2021}, integrating to $Q=1$. More generally, the topological charge density can be expressed as ${\cal Q}=\hat{\bf z}\cdot (\nabla\times {\bf J})$, where the current density for light components $E\customsub{R,L}$ reads~\cite{Volovik1977,Sugic2021,Parmee2022} 
\begin{equation}
    {\bf J} = \frac{1}{2\pi} \frac{{\rm Im} (E^*\customsub{R} \nabla E\customsub{R}+ E^*\customsub{L} \nabla E\customsub{L})}{ |E\customsub{R}|^2 + |E\customsub{L}|^2},
\end{equation}
with the analogous expression for matter obtained by interchanging $E\customsub{R,L} \leftrightarrow\psi_{1,2}$. Here,  ${\bf J}$
represents the gauge potential for the effective magnetic field~\cite{Parmee2022}. In the analogous 3D particle-like skyrmions, the topological charge density translates into the linking number density of ${\bf J}$~\cite{Volovik1977}, similar to the Chern-Simons term for magnetic helicity~\cite{Jackiw00}.

\begin{figure}[t!]
 \centering
\includegraphics[width=0.48\textwidth]{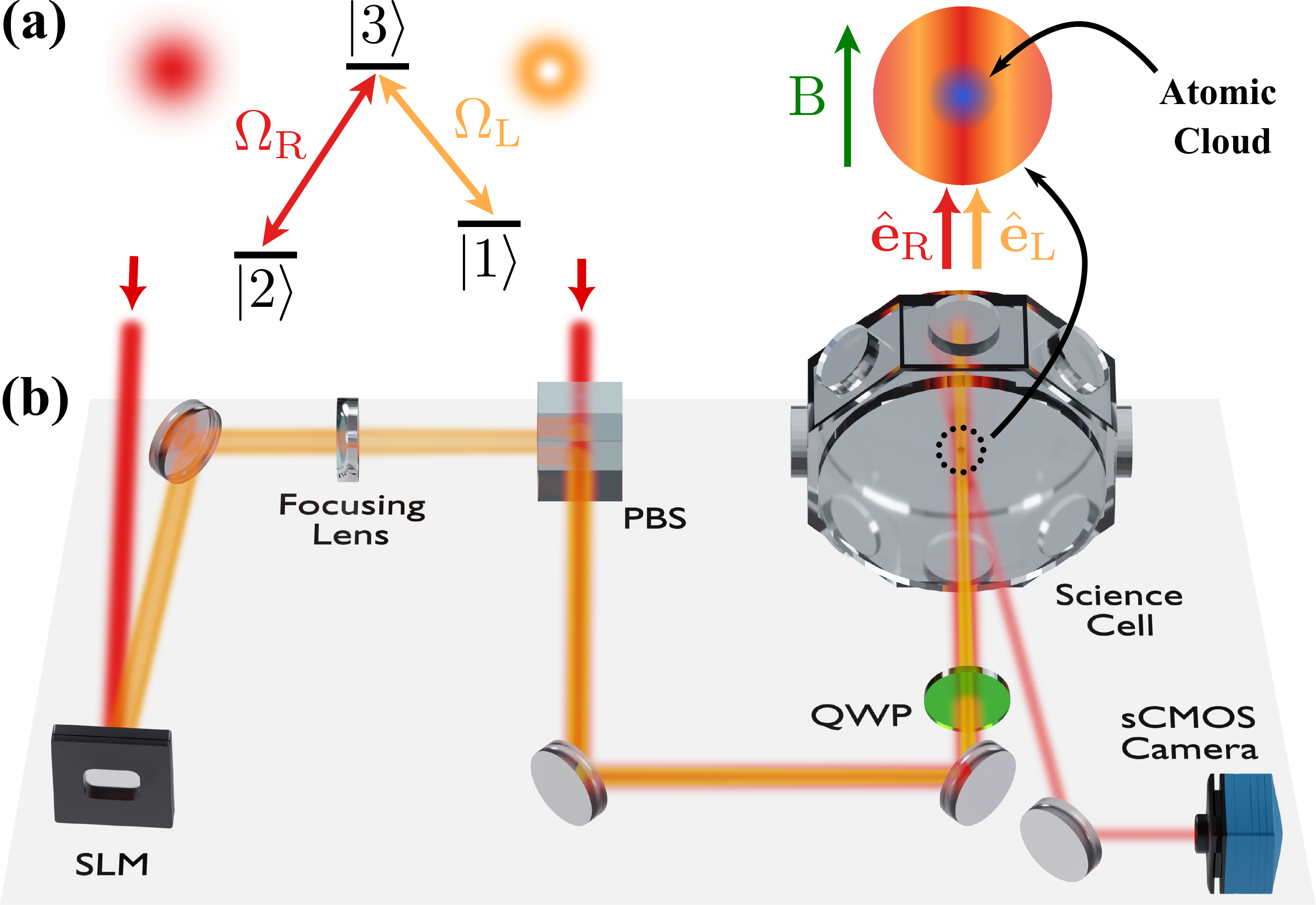}
\caption{{\bf Experimental setup.} {\bf (a)} The atomic-level structure correspond to a $\Lambda$-scheme, with two ground states and one excited state. The Gaussian and Laguerre-Gaussian (LG) beams have opposite circular polarization and are resonant with the $\ket{2}\rightarrow\ket{3}$ and $\ket{1}\rightarrow\ket{3}$ transitions, respectively. {\bf (b)} Schematic of the experimental apparatus. The LG beam (orange) is produced employing a spatial light modulator (SLM) and overlapped with a Gaussian beam (red) of orthogonal linear polarization through a polarizing beam splitter (PBS). Both beams pass through a quarter wave plate (QWP) to acquire the desired polarizations. State-sensitive shadow imaging is performed with a resonant beam (light red) at an angle of $16^\circ$ to extract the spatial profiles of the ground-state populations. 
}
\label{fig:setup}
\end{figure} 

Given the initial state $\Psi_0=[1,0]^T$, the required unitary transformation to achieve the state in Eq.~\eqref{eq:skystate} is represented by $\Psi=U\Psi_0$ where
\begin{equation}
    U= e^{-i\alpha/2} e^{ i \beta \hat{\boldsymbol \rho}\cdot{\boldsymbol \sigma}/2} =  e^{-i\alpha/2}  e^{ i \arctan(\rho) \hat{\boldsymbol \rho}\cdot{\boldsymbol \sigma} } ,
    \label{eq:unitary}
\end{equation}
for coordinates ${\hat{\boldsymbol \rho}} = (x,y,0)/(x^2+y^2)^{1/2}$ on the plane. 
To generate such a transformation by the optical skyrmion field, we consider a gas of cold 87-strontium atoms \ChM{(see Appendix A)} within a $\Lambda$-scheme, as shown in Fig.~\ref{fig:setup}a. The transferred skyrmion within matter is determined by the pseudospin components of two Zeeman ground states $\psi_1\equiv\ket{1}$ and $\psi_2\equiv\ket{2}$. The coupling is induced through the excited level $\ket{3}$, where the LG $E\customsub{L}$ and Gaussian $E\customsub{R}$ beams, with orthogonal circular polarizations $\pol\customsub{L}$ and $\pol\customsub{R}$, respectively, drive the $\ket{1}\rightarrow\ket{3}$ and $\ket{2}\rightarrow\ket{3}$ transitions. The strength of the coupling is characterized by the Rabi frequencies $\Omega\customsub{L}=\mu_{1,3}E\customsub{L}/\hbar$ and $\Omega\customsub{R}=\mu_{2,3}E\customsub{R}/\hbar$, where $\mu_{j,3}$ denote the dipole matrix elements.

We enable the interaction between the optical skyrmion and matter through an adiabatic passage into a dark state~\cite{FleischhauerEtAlRMP2005}. This method is based on resonant interactions, unlike off-resonance imprinting of LG beams~\cite{Andersen06,Wright09}, and can be employed in storing vortices on atomic ensembles~\cite{dutton_prl_2004}. The experimental study of off-resonance imprinting~\cite{Wright09} was extended in Ref.~\cite{Leslie2009} to report the creation of an atomic skyrmion. However, due to a misinterpretation of the $F=2$ spin manifold, the prepared object did not represent a skyrmion~\cite{Leslie_err}, but rather a singular spin-2 vortex~\cite{kawaguchi_physrep_2012}.  

In the dark state passage, the initial state $\psi_1$ adiabatically follows the dark state 
\begin{equation}\label{eq_atom_State}
    \ket{{\rm D}}=- \sin{(\theta/2)} e^{i\phi/2}\ket{1}+  \cos{(\theta/2)} e^{-i\phi/2} \ket{2}.
\end{equation}
Comparing this with Eq.~\eqref{eq:skystate} demonstrates that if the incident field ${\bf E}$ exhibits the skyrmion profile of Eq.~\eqref{eq:skystate}, wrapping once over the Poincar\'e sphere, then the topological texture in Eq.~\eqref{eq_atom_State} represents the same structure if we identify $E\customsub{R}\leftrightarrow \psi_1$ and $E\customsub{L}\leftrightarrow -\psi_2$, with the dark state mixing angle $\tan({\theta/2})=\tan({\theta_o/2})=|E\customsub{L}/E\customsub{R}|$ and the relative phase $\phi=\phi_o={\rm arg} (E\customsub{L}/E\customsub{R})$ corresponding to the order-parameter coordinates on a Poincar\'e sphere. In our scenario, however, the imbalance in the atom-light coupling strength results in a slightly adjusted $\tan({\theta/2})=\tan({\theta_o^c/2})=|\Omega\customsub{L}/\Omega\customsub{R}|$, causing the detected skyrmion in matter to appear slightly skewed compared with the original optical skyrmion. Additionally, we ensure all the atoms are illuminated by choosing the beam width significantly larger than the radius of the atom cloud. Nevertheless, as shown below, the portion of the skyrmion that overlaps with the atom cloud is effectively 
transferred to matter, and revealed in atomic measurements.

A schematic of the experiment is shown in Fig.~\ref{fig:setup}b. To generate the optical skyrmion, a Gaussian beam is split into two orthogonal polarization components, with one component transformed into an LG beam using a spatial light modulator \ChM{(see Appendix B)}. The two polarization components are then recombined to create a beam exhibiting the desired transverse skyrmion pattern. This beam is directed into the cold-atom cloud ($1/e^2$-waist $\rho\customsub{c}\simeq 107(3)\,\mu$m) and resonantly coupled to the atomic transitions. The atomic skyrmion properties are extracted from the state populations measured after various interaction sequences with optical beams \ChM{(see Appendix B)}.

\begin{figure}[t!]
 \centering
 \includegraphics[width=0.48\textwidth]{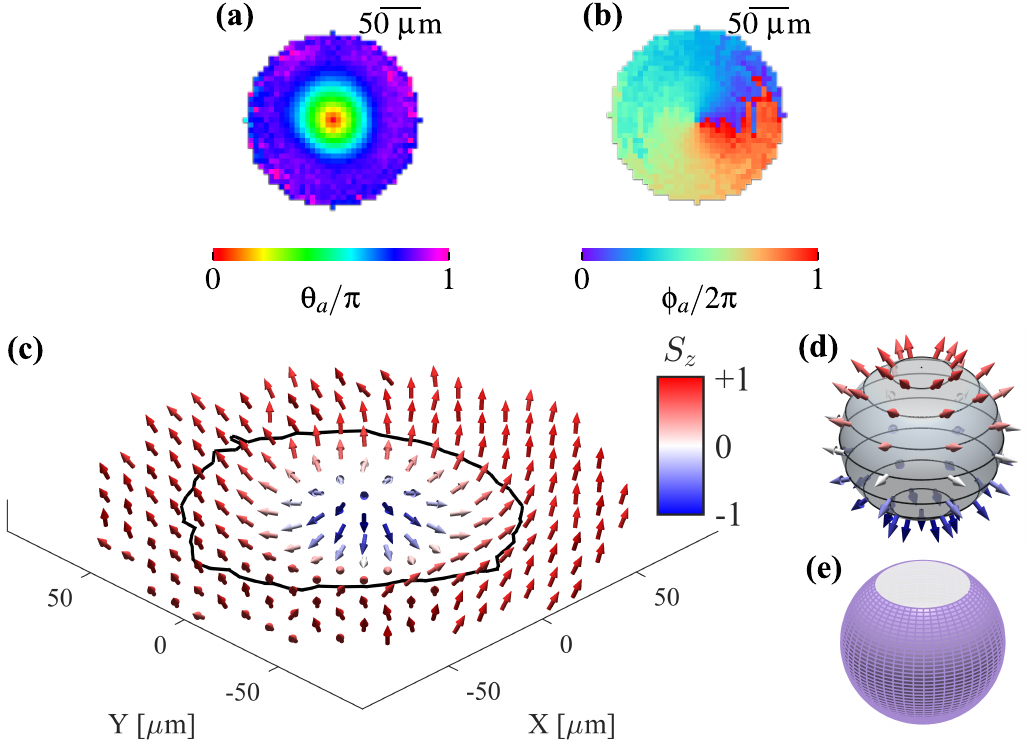}
\caption{\textbf{Dark state reconstruction and skyrmion in matter.} {\bf (a)} Spatial profile of the dark state mixing angle $\theta\customsub{a}$ and {\bf (b)} azimuthal angle $\phi\customsub{a}$ [see Eq.~\eqref{eq_atom_State}], extracted from state-population measurements. Images on panels (a) and (b) are averaged over 100 experimental runs \ChM{(see Appendix B)}. \textbf{(c)} Pseudospin texture of the transferred skyrmion on the $xy$ plane, determined by the atomic dark state population and coherence. The curve determines the path along which the spin vector maintains a constant
angle 0.8$\pi$ between the radial and vertical axes. \textbf{(d)} The region
outlined by this path is mapped onto a sphere, analogously to the optical skyrmion in Fig.~\ref{fig:optical_Skyrmion_texture}. \textbf{(e)} Mapping of the spin texture on the $S^2$ order-parameter space, as determined by the Poincar\'e sphere, nearly wraps over the entire sphere, resulting in the topological charge of $Q \simeq 0.84$.}
 \label{fig:atomic_Skyrmion_texture}
\end{figure}

In the dark state adiabatic passage of the optical skyrmion to matter, the atoms are initially optically pumped into the $\ket{1}$ state. The Gaussian beam on the $\ket{2}\rightarrow\ket{3}$ transition is subsequently turned on, followed by an adiabatic ramp of the LG beam on the $\ket{1}\rightarrow\ket{3}$ transition. This process adiabatically connects the initial $\ket{1}$ state to the dark state of Eq.~\eqref{eq_atom_State}. 
By assuming that the atoms are in the dark state of Eq.~\eqref{eq_atom_State}, we then determine from the measurement data the unknown parameters $\theta$ and $\phi$, subsequently referred to as $\theta_a$ and $\phi_a$ \ChM{(see Appendix B)}. 
The resulting profiles are shown in Figs.~\ref{fig:atomic_Skyrmion_texture}a,b.

\begin{figure}[t!]
 \centering
 \includegraphics[width=0.48\textwidth]{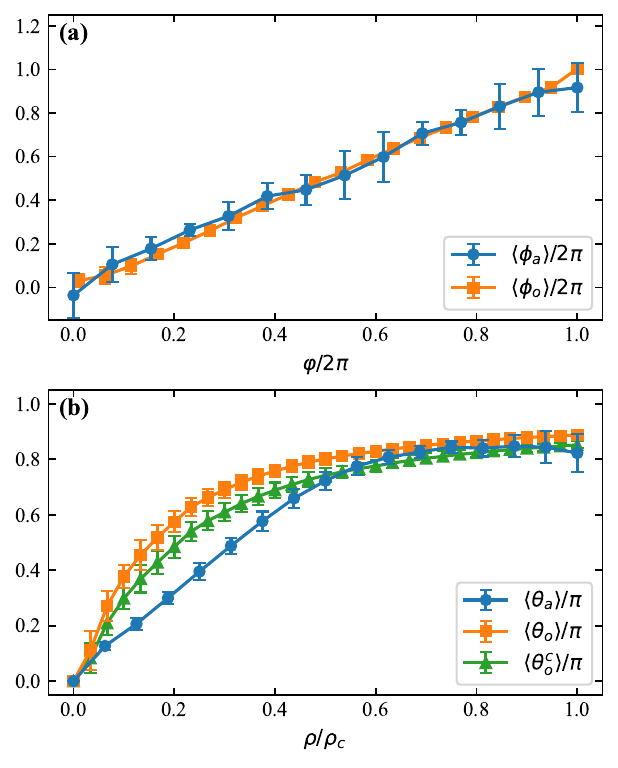}
 \caption{
 \textbf{Efficiency of the topological skyrmion transfer from light to matter.} {\bf (a)} \ChM{Radially averaged spin azimuthal angle $\langle\phi_a\rangle$ (blue circles) for the atomic and $\langle\phi_o\rangle$ (orange squares) for the optical skyrmion, as a function of the azimuthal coordinate $\varphi$ [see definitions in and around Eq.~\eqref{eq_atom_State}].} The error bars represent the standard deviation of the \ChM{spin} azimuthal angle using radial sampling. {\bf (b)} \ChM{Azimuthally averaged mixing angle $\langle\theta_a\rangle$ (blue circles) for the atomic and $\langle\theta_o\rangle$ (orange squares) for the optical skyrmion, as a function of the radial coordinate $\rho$.} The error bars represent the standard deviation of the \ChM{mixing} angle using azimuthal sampling. The unbalanced light-matter coupling strength results in a corrected ideal transferred mixing angle $\langle\theta^c_o\rangle$ (green triangles), with an algebraic dependence on $\langle\theta_o\rangle$ \ChM{(see Appendix A)}. 
}
  \label{fig:Skyrmion_parameter_transfer}
\end{figure}

The pseudospin texture of the transferred skyrmion is shown in Fig.~\ref{fig:atomic_Skyrmion_texture}c on the $xy$ plane, along with its compactified version on a sphere (Fig.~\ref{fig:atomic_Skyrmion_texture}d). We also display the
Poincar\'e sphere representation, illustrating how the Stokes vectors cover most of the order-parameter space (Fig.~\ref{fig:atomic_Skyrmion_texture}e), defined by the angles ($0\leq\theta_a\leq\pi,0\leq\phi_a\leq2\pi$) in Eq.~\eqref{eq_atom_State}. Integrating the topological charge from the experimental data according to Eq.~\eqref{Eq:SkyrmionNumber} yields $Q \simeq 0.84$.

In Fig.~\ref{fig:Skyrmion_parameter_transfer}, we show radially averaged relative phases and azimuthally averaged mixing angles, representing the Poincar\'e sphere spherical angles, for the different scenarios of Eq.~\eqref{eq_atom_State}, when the optical and matter skyrmions are determined within the same spatial region. Given the imbalance in the atom-light coupling strengths between the Gaussian and LG beams \ChM{(see Appendix A)}, an ideal transfer of topology would yield $\phi\customsub{a}=\phi\customsub{o}$ and $\theta\customsub{a}=\theta\customsub{o}^c$, as explained previously. The former equality is experimentally well-verified, while some mismatches are observed in the latter, likely due to residual aberrations in our imaging system.

\section{Discussion}

We demonstrated that topological optical skyrmion textures can be effectively mapped onto matter by leveraging well-established methods for imprinting optical vortices of LG beams onto atomic vortices~\cite{Andersen06,Pugatch07,Wright09,Moretti09}.
Due to the smaller spatial region of the atom cloud, the spin vector orientation between the radial and vertical axes at the edge of the cloud maintains a smaller value $0.8\pi$ than in the optical skyrmion. When integrating over the same spatial area, the optical skyrmion's topological charge is 0.84, which decreases to 0.80 when accounting for the imbalance in atom-light coupling strength. This value is in reasonable agreement with the atomic skyrmion's topological charge of 0.84, with the remaining discrepancy likely due to systematic errors, such as aberrations in our imaging system.

Mapping skyrmion topology onto matter and detecting it through atomic measurements opens up possibilities for topological data encoding and storage. Future work could extend this approach to skyrmion laser pulses, leveraging existing technologies for storing and releasing entire pulses within an atomic cloud~\cite{LiuEtAlNature2001,dutton_prl_2004}. Additionally, our approach offers improved detection and analysis of more complex topologies in structured light.  This is particularly relevant because the traditional description of paraxial light waves in terms of the Stokes vectors on a Poincar\'e sphere is incomplete; it discards the variation of the total electromagnetic phase of vibration~\cite{Sugic2021,Parmee2022}. At optical frequencies, the spatial profile of this phase is challenging to measure due to rapid oscillations~\cite{Sugic2021} and is typically overlooked in the topological analyses. However, this phase variation leads to an optical hypersphere description and the emergence of genuinely 3D topologies. The most dramatic examples are full 3D particle-like skyrmions and hopfions~\cite{Sugic2021,Parmee2022}, analogous to knotted solitons of the Skyrme-Faddeev model in high-energy physics~\cite{faddeev_nature_1997,manton-sutcliffe}. Such constructions based on links and knots are related to fundamental ideas proposed by Kelvin, who suggested them as building blocks of atomic particles~\cite{Thomson1869}. In the context of electromagnetic fields, they have been suggested even as models of ball lightning~\cite{Ranada1996,Ranada1998}. Once transferred to atomic media,  the detection of these structures no longer faces similar challenges as in the optical case~\cite{Lee2018}.

\begin{acknowledgments}
This work was supported by the CQT/MoE (Grant No.\ R-710-002-016-271), the Singapore Ministry of Education Academic Research Fund Tier2 (Grant No.\ MOE-T2EP50220-0008) and EPSRC UK (Grant No.\ EP/S002952/1).
\end{acknowledgments}

Datasets will be provided in a data repository (doi: to be added later).

\appendix

\section{Atomic sample}

The optical skyrmion is coupled to an cold 87-strontium gas at a temperature of $6.9(3)\,\mu$K, using the $\ket{\,^1S_0, F=9/2}\rightarrow\ket{\,^3P_1, F=9/2}$ hyperfine transition of the intercombination line at $689\,$nm \cite{yang2015high}. The gas contains $4.2(3)\times10^6$ atoms, within a $1/e^2$-width $\rho\customsub{c}=107(3)\,\mu$m of a quasi-Gaussian shape. A $\Lambda$-scheme connects the Zeeman ground states $\psi_1\equiv\ket{1}\equiv\ket{m=9/2}$ and $\psi_2\equiv\ket{2}\equiv\ket{m=5/2}$ to the excited state $\ket{3}\equiv\ket{m=7/2}$. The LG and Gaussian beam polarizations and frequencies are set to couple to the $\ket{1}\rightarrow\ket{3}$ and $\ket{2}\rightarrow\ket{3}$ transitions, respectively. 
The Clebsch-Gordan coefficients  $C_1=-\sqrt{18/99}$ and $C_2=\sqrt{32/99}$, associated with the atom-light coupling strengths of these transitions, differ, resulting in unbalanced dipole matrix elements $\mu_{i,3}=C_i\mu$, where $\mu$ denotes the reduced dipole moment. This imbalance affects the radial mixing angle $\theta_o$, yielding the corrected value $\theta^c_o$, such that $\tan({\theta^c_o/2})/\tan({\theta_o/2})=(18/32)^{1/2}$ in Fig.~\ref{fig:Skyrmion_parameter_transfer}b.

\ChM{The skyrmion lifetime in this experiment is mainly limited by the thermal motion of the atoms which scrambles the topological structure. The lifetime can be estimated to be on the order of $\tau\simeq d/\bar{v}$, where $\bar{v}\simeq 2.5\,\text{cm/s}$ is the thermal speed, and $d\simeq 5\,\mu$m the typical size of the topological structure, leading to $\tau\simeq 200\,\mu$s. For applications with long storage times, significantly longer skyrmion lifetimes can be achieved with atomic Bose-Einstein condensates (in rotation, only limited by the condensate lifetime up to tens of seconds) or in optical lattice potentials that freeze the atomic motion.}

\section{Measurements of optical and atomic amplitude and phase profiles}

An optical skyrmion is represented by a Poincar\'e beam~\cite{Beckley2010}, generated by a superposition of a Gaussian and an LG beam with orthogonal polarizations~\cite{Donati2016,Gao2020}. To produce the LG beam, a Gaussian beam is first split into two orthogonal polarization components. One component is transformed into an LG beam using a phase mask imprinted on a spatial light modulator, employing a digital-hologram method \cite{Arrizon07}. To measure the relative phase profile of the Gaussian and the LG beams, we use a polarizer to project the polarization of each beam. The resulting interference pattern is detected on a CCD camera. 

To characterize the atomic skyrmion, the populations of states $\ket{1}$ and $\ket{2}$ are detected using spin-sensitive shadow imaging techniques~\cite{leroux2018non}. More precisely, the $\ket{1}$ population profile is measured with a circularly polarized probe beam resonantly tuned to the $\ket{\,^1S_0, F=9/2, m=9/2}\rightarrow\ket{\,^3P_1, F=11/2, m=11/2}$ transition. Subsequently, the $\ket{2}$ population profile is measured by tuning the probe beam to the $\ket{\,^1S_0, F=9/2, m=5/2}\rightarrow\ket{\,^3P_1, F=11/2, m=7/2}$ transition. 
The $\theta_a$ mixing angle profile is extracted from population measurements following the adiabatic transfer of the optical skyrmion to the atomic ensemble. The relative phase $\phi\customsub{a}$ is extracted after the transfer by switching off the LG beam, and simultaneously turning on an equally polarized co-propagating plane wave that addresses the same transition. If the relative local phase between the two beams is zero with equal field amplitude, the atoms remain in the dark state defined by Eq.~(\ref{eq_atom_State}). Otherwise, a fraction of the $\ket{1}$ and $\ket{2}$ populations will be excited to the state $\ket{3}$, leading to an asymmetric profile from which $\phi\customsub{a}$ can be extracted. 

As beams follow different optical paths before being recombined, their relative phase is not stable, resulting in shot-to-shot random variations. However, the relative phase, with a coherence time of about $10\,$ms, does not significantly vary during the short interrogation time of about $~100\,\mu$s (duty cycle $\sim 0.01\%$). If an image, obtained after each run, contains phase information, it is fitted and properly reoriented to avoid phase blurring during the summation of the images. The latter is used to improve the signal-to-noise ratio.

\bibliography{atomlightFirstNameTitle,Bib_Skyrmion}

\end{document}